\documentstyle[eqsecnum,aps,epsf,amsmath]{revtex}
\begin{document}

\draft

\newcommand {\beq}{\begin{eqnarray}}
\newcommand {\eeq}{\end{eqnarray}}
\newcommand {\be}{\begin{equation}}
\newcommand {\ee}{\end{equation}}
\newcommand{\NN}{\nonumber \\}
\newcommand{\Gmu}{\gamma^{\mu}}
\newcommand{\Gnu}{\gamma^{\nu}}
\newcommand{\gmu}{\gamma_{\mu}}
\newcommand{\gnu}{\gamma_{\nu}}
\newcommand{\bg}{\mbox{\boldmath $\gamma$}}
\newcommand{\gfour}{\gamma_4}
\newcommand{\del}{\partial}
\newcommand{\k}{\mbox{\boldmath $k$}}
\newcommand{\q}{\mbox{\boldmath $q$}}
\newcommand{\p}{\mbox{\boldmath $p$}}
\newcommand{\x}{\mbox{\boldmath $x$}}
\newcommand{\y}{\mbox{\boldmath $y$}}
\newcommand{\X}{\mbox{\boldmath $X$}}
\newcommand{\wk}{\omega_{\mbox{\boldmath \scriptsize $k$}}}
\newcommand{\wm}{\omega_{\mbox{\boldmath \scriptsize $p$}}}
\newcommand{\wone}{\omega_{\mbox{\boldmath \scriptsize $k$}_1}}
\newcommand{\wtwo}{\omega_{\mbox{\boldmath \scriptsize $k$}_2}}
\newcommand{\wthree}{\omega_{\mbox{\boldmath \scriptsize $k$}_3}}
\newcommand{\wmin}{\omega_{\mbox{\boldmath \scriptsize $p$} -\mbox{\boldmath \scriptsize $k$}_1-\mbox{\boldmath \scriptsize $k$}_2 }}
\newcommand{\La}{{\Lambda_{\scriptsize{\mbox{QCD}}}}}
\newcommand{\CC}{\langle \bar{q} q \rangle}
\newcommand{\ltgt}{^{\genfrac{}{}{0pt}{}{<}{>}}}

\title{
The effect of memory on relaxation in a scalar field theory
}
 
\author{
Takashi Ikeda \footnote{\tt Email: ikeda@bnl.gov}
}

\address{
RIKEN BNL Research Center, Brookhaven
National Laboratory, Upton, New York 11973-5000, USA
}

\maketitle

\begin{abstract}

 We derive a kinetic equation with a non-Markovian collision term which
 includes a memory effect, from Kadanoff-Baym equations in $\phi^4$
 theory within the three-loop level for the two-particle
 irreducible (2PI) effective action.
 The memory effect is incorporated
 into the kinetic equation by a generalized Kadanoff-Baym ansatz.
 Based on the kinetic equations with and without the memory effect,
 we investigate an influence of this effect on
 decay of a single particle excitation with zero momentum
 in 3+1 dimensions and the spatially homogeneous case. 
 Numerical results show that, while the time evolution of the zero mode
 is completely unaffected by the memory effect due to a separation of
 scales in the weak coupling regime, this effect leads first to faster
 relaxation than the case without it and then to slower relaxation as
 the coupling constant increases. 

\end{abstract}

\maketitle


\section{Introduction}

 The vacuum of the quantum chromodynamics (QCD) is believed to
undergo a phase transition from the hadronic phase
to the quark-gluon plasma (QGP)
at high temperature $T$ and/or at high quark chemical potential $\mu$,
and such a new state of matter is expected
to be produced in heavy-ion collision experiments ongoing
at Relativistic Heavy-Ion Collider (RHIC)
and in the future at Large Hadron Collider (LHC)
\cite{QMproceedings}. 
Motivated by these experiments as well as 
the inflationary dynamics in the early universe
\cite{Son:1996uv,Khlebnikov,Micha:2002ey,Felder:2000hq,Felder:2000hr}, 
quantum field theories
in and out of thermal equilibrium have been extensively studied. 
Many theoretical explanations related to the experiments
have been made on the basis of thermal and chemical
equilibrium. Here, there is an open question about justifying that QGP
produced in a transient state achieve the equilibration before
the phase transition to the hadronic matter.
Thermalization of QGP from the initial state away from thermal equilibrium 
\cite{MVmodel} 
has been studied on the basis of the classical Boltzmann equations
\cite{Mueller,Bottomup,Muller,Arnold:2003rq}. 
However, it has not been established yet.
It might be important to understand the early stages in heavy-ion
collisions from the aspect of nonequilibrium phenomena, 
compared to a thermally equilibrated state having no information about 
the initial conditions.

The classical Boltzmann equations are often used for studying nonequilibrium
properties. Their collision term is {\it Markovian}
one which treats scatterings between particles as independent ones, and 
no past information is needed to calculate it.
As a consequence, they cannot be applicable to the strongly
correlated or dense plasmas. Undoubtedly, {\it generalized} kinetic
equations are necessary in order to overcome their limitations.

The quantum Kadanoff-Baym (KB) equations describe the time evolution of
the two-time Green's functions which involve the information about both 
the dynamical spectral function and the one-particle distribution
function in the correlated quantum plasmas
\cite{KBbook}, 
and are designed to study time-dependent nonequilibrium phenomena
\cite{Aarts:2001qa,Berges:2000ur,Berges:2001fi,Berges,Cooper:2002qd,Berges:2002wr,Cassing,Boyanovsky:1996xx,Boyanovsky:1997zg,Boyanovsky:1999cy,Boyanovsky:2003tc,Aarts:2000mg,Salle,Ikezi:2003kn}.
In general, it is difficult to solve the KB equations directly
due to two-time correlation. It is easier to analyze
kinetic equations for Wigner distribution function $N$ which can be
derived from the KB equations in the equal-time limit, by imposing some
approximations. When deriving the kinetic equations, 
one encounters the reconstruction problem: 
How can two-point Green's functions be expressed as a function of $N$ ? 
While the simplest solution is the Kadanoff-Baym (KB) ansatz
which is mostly used for deriving Boltzmann-like equations,
we have a more general solution, the generalized Kadanoff-Baym (GKB) 
ansatz proposed by Lipavsky {\it et al.} 
\cite{Lipavsky}. 
Although the KB ansatz leads to inconsistencies in time arguments of a
{\it memory effect} in the collision term, the GKB ansatz is able to 
take into account it properly, as we will see below.
This effect is expected to play an important role in correlated 
quantum systems and/or short-time phenomena.

The purpose of this paper is to study an influence of the memory
effect on the time evolution of the distribution function, and to
point out that relaxation is strongly affected by the this effect,
on the the basis of the kinetic equations.
Our starting point is a simple question: 
`` If successive collisions among particles in a system of interest 
are not independent ones unlike in the Boltzmann equations, how
does relaxation of the system be affected ? ''
A {\it non-Markovian extension} of the Boltzmann collision term 
is derived by using the GKB ansatz,
and one has to perform a temporal integration over the distribution
function having past times in it, as in the original KB equations. 
However, computational efforts are reduced remarkably 
compared to the KB equations because the distribution function has 
to be known and time evolved only along the one-time direction.
On the other hand, in the KB equations,
Green's functions have two time arguments 
and time evolved in the two-time plane. 
Usually, one expects that, in the weak coupling regime
the relaxation time can be evaluated correctly by the 
Boltzmann equations due to a separation of scales.
However, as we will see, there is the onset of the memory effect 
in the region where the separation of scales is satisfied.

The present paper is organized as follows. 
In Sec. II, the
KB equations for $\phi^4$ theory within the 3-loop level for 2PI effective
action, and the KB and GKB ansatz as approximate solutions of the
reconstruction problem are reviewed. 
Then, with the GKB ansatz, non-Markovian and Markovian collision
terms are constructed, and kinetic equations with and without the memory 
effect are derived. Further,
those are reduced to linear equations for the distribution
function in the case of a single particle excitation. 
Section III deals with numerical calculations. After renormalization by
the modified MS scheme in a gap equation, 
the relation between the thermal mass and the coupling
constant is determined self-consistently by solving the gap equation
numerically. With the solutions of the gap equation, the time-dependent
damping rate with zero momentum 
which is a key quantity in the kinetic equations is calculated
for several values of the coupling constant. Then, the time
evolutions of the zero mode are obtained from
the kinetic equations with/without the memory effect.
By changing the coupling constant, an influence of the memory effect 
on the relaxation of the excitation is examined.
Some discussions about the numerical results are given at the end of
this section.
Section IV is devoted to summary and outlook.

\section{From Kadanoff-Baym equations to Kinetic equations}

KB equations are designed to study nonequilibrium
phenomena of a quantum system in terms of one-body Green's functions. 
In this section, we study how kinetic equations are derived 
from KB equations, by imposing some approximations.
We will see that the memory effect which is neglected in the classical 
Boltzmann equations is incorporated into kinetic
equations by the GKB ansatz.

\subsection{KB equations for $\phi^4$ theory}

We consider a simple scalar field theory with a Lagrangian density in 3+1
space-time dimensions,
\be
{ \cal L} = \frac{1}{2}\del_{\mu}\phi(x)\del^{\mu}\phi(x) -
\frac{m^2}{2}\phi^2(x) + \frac{\lambda}{4!}\phi^4 (x) \, ,
\label{eq:lagrangian}
\ee
where $x = (x_0,\x)$, $m$ is the bare mass and $\lambda$ denotes the
bare coupling strength. This is an extensive 
example of a relativistic quantum field theory which allows us to 
challenge theoretical calculations without encountering troubles of 
gauge invariance.

\begin{figure}[htbp]
    \centerline{
      \epsfxsize=0.49\textwidth
      \epsfbox{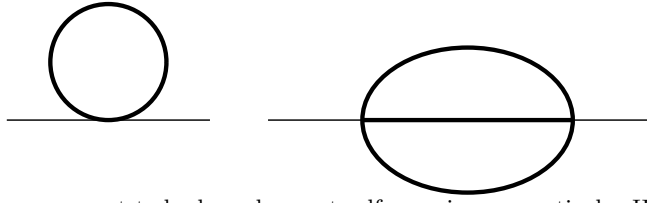}
    }
  \caption{ Left and right diagrams represent tadpole and sunset
      self-energies, respectively. Heavy lines denote dressed propagators.
          } 
  \label{fig:selfenergy}
\end{figure}

Within the three-loop order for 2PI effective action, 
KB equation for $G^<$ on Schwinger-Keldysh contour 
\cite{KBbook,Schwinger:1960qe,Keldysh:ud}
takes a following form in the spatially homogeneous case
\cite{Aarts:2001qa,Cassing,Blaizot:2001nr},
\beq
\left( \del_{x_0}^2 + \p^2 + m^2 + \Sigma^{\delta}(x_0,x_0) \right) 
G^<(\p,x_0,y_0)
&=& -i \int_{t_0}^{x_0}dz_0
\left[\Sigma^>(\p,x_0,z_0)-\Sigma^<(\p,x_0,z_0)\right] G^<(\p,z_0,y_0)
\nonumber \\
& & +i \int_{t_0}^{y_0}dz_0 \Sigma^<(\p,x_0,z_0)
\left[ G^>(\p,z_0,y_0)- G^<(\p,z_0,y_0) \right] \, ,
\label{eq:KBforGp}
\eeq
where $\Sigma^{\delta}(x_0,x_0)$ 
is a tadpole self-energy in the leading order of the coupling constant 
and $\Sigma^{\genfrac{}{}{0pt}{}{<}{>}}(\p,x_0,y_0)$ are 
sunset self-energies in the next-to leading order, and 
corresponding diagrams are shown in 
Fig. \ref{fig:selfenergy}. These self-energies
are expressed in terms of $G^{\genfrac{}{}{0pt}{}{<}{>}}$ as
\beq
\Sigma^{\delta}(x_0,x_0) &=& \frac{\lambda}{2} \int \frac{d^3 k}{(2\pi)^3} 
G^<(\k,x_0,x_0) \, ,
\label{eq:tadpole_in_p} \\
\Sigma^{\genfrac{}{}{0pt}{}{<}{>}}(\p,x_0,y_0) &=& - \frac{\lambda^2}{6}
\int \frac{d^3 k_1}{(2\pi)^3} \frac{d^3 k_2}{(2\pi)^3} \frac{d^3
k_3}{(2\pi)^3} (2\pi)^3 \delta^{(3)}(\k_1+\k_2+\k_3-\p) \nonumber \\
&& \times
G^{\genfrac{}{}{0pt}{}{<}{>}}(\k_1,x_0,y_0)
G^{\genfrac{}{}{0pt}{}{<}{>}}(\k_2,x_0,y_0)
G^{\genfrac{}{}{0pt}{}{<}{>}}(\k_3,x_0,y_0)
\, .
\label{eq:sunset_in_p}
\eeq
Since $G\ltgt$ are full Green's functions, the self-energies are
determined self-consistently. 
We shall discuss this later.

Although one can analyze the KB equations directly,
it is easier to consider kinetic equations for the 
Wigner distribution function, which we are concerned with here.
For this purpose, it is useful to observe the KB equation (\ref{eq:KBforGp}) 
in the equal-time limit $x_0=y_0=t$,
\be
\left( \del_{x_0}^2 + \p^2 + M^2(x_0) \right) 
G^<(\p,x_0,y_0)|_{x_0=y_0=t} 
= -i \int_{t_0}^{t}dz_0
\left[\Sigma^>(\p,t,z_0)G^<(\p,z_0,t) - 
\Sigma^<(\p,x_0,z_0) G^>(\p,z_0,y_0)\right] \, , 
\label{eq:ETforGp} 
\ee
where $M^2(x_0) \equiv m^2 + \Sigma^{\delta}(x_0,x_0)$.
Considering the similar equation in which the temporal derivative acts
on $y_0$ and taking the difference of the two equations, 
we obtained the following equation
\cite{Blaizot:2001nr},
\beq
\left( \del_{x_0}^2 - \del_{y_0}^2 \right) G^<(\p,x_0,y_0)|
_{x_0=y_0=t} 
= -i \int_{t_0}^{t} dz_0 & & \left[ \Sigma^>(\p,t,z_0) G^<(\p,z_0,t) \right.
 - \Sigma^<(\p,t,z_0) G^>(\p,z_0,t)
\nonumber \\
& & - G^>(\p,t,z_0) \Sigma^<(\p,z_0,t)
\left.  + G^<(\p,t,z_0) \Sigma^>(\p,z_0,t) \right] \, .
\label{eq:KB_equaltime}
\eeq

\subsection{Generalized Kadanoff-Baym ansatz}

In order to derive the kinetic equation satisfied by the Winger
distribution function $N(\p,t)$ from the KB equation
(\ref{eq:KB_equaltime}), it is needed to
express $G\ltgt (\p,x_0,y_0)$ as a function of $N(\p,t)$. 
This is the so-called reconstruction problem. 
The simplest solution of this problem is the common 
Kadanoff-Baym (KB) ansatz
\cite{KBbook}, given by
\beq
G^<(\k,x_0,y_0) &=& \frac{1}{2\wk(t)} \left[ e^{-i\wk(t) s_0} N(\k,t) 
+ e^{i\wk(t) s_0} \left(1+ N(-\k,t) \right)\right] \, ,
\nonumber \\
G^>(\k,x_0,y_0) &=& \frac{1}{2\wk(t)} \left[ e^{-i\wk(t) s_0} 
\left( 1 + N(\k,t) \right)
+ e^{i\wk(t) s_0} N(-\k,t)\right] \, ,
\label{eq:KB_QPA}
\eeq
with $s_0 = x_0-y_0$, $t = (x_0+y_0)/2$, and 
$\wk(t) = \sqrt{\k^2 + M^2(t)}$.
Here, we have adopted the quasiparticle approximation and
decomposed components of the distribution function with 
the positive and negative energy.
However, this ansatz leads to inconsistencies 
in the time argument of the memory effect because 
the distribution function depends only on the central time $t$.

As a more general ansatz,
a generalized Kadanoff-Baym (GKB) ansatz has been proposed by 
Lipavsky {\it et al.} \cite{Lipavsky}, which is given by 
\beq
G^<(\k,x_0,y_0) &=& \frac{\theta(s_0)}{2 \wk(y_0)}
\left[ e^{-i \wk(y_0) s_0} N(\k,y_0) +
e^{i \wk(y_0) s_0} \left(1+N(-\k,y_0) \right)\right] \nonumber \\
&& + \frac{\theta(-s_0)}{2 \wk(x_0)}\left[ e^{-i \wk(x_0) s_0} N(\k,x_0) +
e^{i \wk(x_0) s_0} \left(1+N(-\k,x_0) \right)\right] \, ,
\nonumber \\
G^<(\k,x_0,y_0) &=& \frac{\theta(s_0)}{2 \wk(y_0)}
\left[ e^{-i \wk(y_0) s_0} \left(1+N(\k,y_0)\right) +
e^{i \wk(y_0) s_0} N(-\k,y_0)\right] \nonumber \\
&& + \frac{\theta(-s_0)}{2 \wk(x_0)}\left[ e^{-i \wk(x_0) s_0} 
\left(1+N(\k,x_0)\right) +
e^{i \wk(x_0) s_0} N(-\k,x_0)\right] \, ,
\label{eq:GKB_QPA}
\eeq
within the quasiparticle approximation.
Note that the GKB ansatz differs from the KB ansatz (\ref{eq:KB_QPA})
in the time argument of distribution functions.
This is why the GKB ansatz is able to takes into account the
memory effect properly.
Therefore, we will use the GKB ansatz below.

\subsection{Kinetic equation with/without the memory effect}

In the GKB ansatz (\ref{eq:GKB_QPA}),
the KB equation in the equal-time limit
(\ref{eq:KB_equaltime}) for the
positive energy component $N(\p,t)$ reduces to 
\beq
&& \frac{\del}{\del t} N(\p,t) =
- \frac{\lambda^2}{6} \int \frac{d^3 k_1}{(2\pi)^3}
\frac{d^3 k_2}{(2\pi)^3} \int_{t_0}^{t} dz_0 
\frac{1}{8\wone \wtwo \wmin \wm}
\nonumber \\
&&\times \left\{ \cos(\wone+\wtwo+\wmin-\wm)(t-z_0)
\left[\bar{N}_{\k_1}\bar{N}_{\k_2}\bar{N}_{\p-\k_1-\k_2} N_{\p}
- N_{\k_1}N_{\k_2}N_{\p-\k_1-\k_2} \bar{N}_{\p}\right]\right.
\nonumber \\
&&\hspace{0.3cm}+ \cos(\wone+\wtwo+\wmin+\wm)(t-z_0)
\left[ N_{-\k_1}N_{-\k_2}N_{\k_1+\k_2-\p} N_{\p}
- \bar{N}_{-\k_1}\bar{N}_{-\k_2}\bar{N}_{\k_1+\k_2-\p}\bar{N}_{\p}\right]
\nonumber \\
&&\hspace{0.3cm}+ 3 \cos(\wone+\wtwo-\wmin-\wm)(t-z_0)
\left[\bar{N}_{\k_1}\bar{N}_{\k_2}N_{\k_1+\k_2-\p} N_{\p}
- N_{\k_1}N_{\k_2}\bar{N}_{\k_1+\k_2-\p} \bar{N}_{\p}\right]
\nonumber \\
&&\hspace{0.3cm}+ 3 \cos(\wone-\wtwo-\wmin-\wm)(t-z_0)
\left. \left[\bar{N}_{\k_1}N_{-\k_2}N_{\k_1+\k_2-\p} N_{\p}
- N_{\k_1}\bar{N}_{-\k_2}\bar{N}_{\k_1+\k_2-\p} \bar{N}_{\p}\right]
\right\}
\, ,
\label{eq:nonMarkov}
\eeq
where $\wk \equiv \wk(z_0)$, $N_{\k} \equiv N(\k,z_0)$ and 
$\bar{N}_{\k} \equiv 1 + N(\k,z_0)$.
The right-hand side of (\ref{eq:nonMarkov})
is the non-Markovian extension of the standard Boltzmann collision term.
The memory effect is described by the integration over past times
which the distribution functions have. This effect is
expected to be evidently important if the relaxation time of the system
of interest is shorter than the memory time itself, and its importance has
been illustrated in various non-relativistic systems
\cite{Kohler,Kohler2}. 
For the KB ansatz, the time argument of distribution functions 
and energies in the integral would have
been $(z_0+t)/2$ instead of $z_0$, which is inconsistent with the
time argument of the memory effect as we mentioned in the introduction.

If neglecting the memory effect, that is, $N(\k,z_0) \to N(\k,t)$ 
and $\wk(z_0) \to \wk(t)$ in the collision term,
the integration over $z_0$ can be performed analytically, and we obtain 
the finite-time generalization of the
Boltzmann equation with the Markovian collision term,
\beq
&& \frac{\del}{\del t} N(\p,t) =
- \frac{\lambda^2}{6 \wm} \int \frac{d^3 k_1}{(2\pi)^3}
\frac{d^3 k_2}{(2\pi)^3} \frac{1}{8\wone \wtwo \wmin}
\nonumber \\
&&\times \left\{ 
\frac{\sin(\wone+\wtwo+\wmin-\wm)(t-t_0)}{\wone+\wtwo+\wmin-\wm}
\left[\bar{N}_{\k_1}\bar{N}_{\k_2}\bar{N}_{\p-\k_1-\k_2} N_{\p}
- N_{\k_1}N_{\k_2}N_{\p-\k_1-\k_2} \bar{N}_{\p}\right]\right.
\nonumber \\
&&\hspace{0.3cm}+ 
\frac{\sin(\wone+\wtwo+\wmin+\wm)(t-t_0)}{\wone+\wtwo+\wmin+\wm}
\left[ N_{-\k_1}N_{-\k_2}N_{\k_1+\k_2-\p} N_{\p}
- \bar{N}_{-\k_1}\bar{N}_{-\k_2}\bar{N}_{\k_1+\k_2-\p}\bar{N}_{\p}\right]
\nonumber \\
&&\hspace{0.3cm}+ 3 
\frac{\sin(\wone+\wtwo-\wmin-\wm)(t-t_0)}{\wone+\wtwo-\wmin-\wm}
\left[\bar{N}_{\k_1}\bar{N}_{\k_2}N_{\k_1+\k_2-\p} N_{\p}
- N_{\k_1}N_{\k_2}\bar{N}_{\k_1+\k_2-\p} \bar{N}_{\p}\right]
\nonumber \\
&&\hspace{0.3cm}+ 3 
\frac{\sin(\wone-\wtwo-\wmin-\wm)(t-t_0)}{\wone-\wtwo-\wmin-\wm}
\left. \left[\bar{N}_{\k_1}N_{-\k_2}N_{\k_1+\k_2-\p} N_{\p}
- N_{\k_1}\bar{N}_{-\k_2}\bar{N}_{\k_1+\k_2-\p} \bar{N}_{\p}\right]
\right\}
\, .
\label{eq:Markov}
\eeq
Here, $\wk \equiv \wk(t)$, 
$N_{\k} \equiv N(\k,t)$ and  $\bar{N}_{\k} \equiv 1 + N(\k,t)$.

It is instructive to notice that,
by letting the finite initial time $t_0 \to -\infty$ 
corresponding to the limit of complete collisions
and using the identity
$ \sin(xt)/x \xrightarrow[t \to \infty]{} \pi \delta(x)$, 
Eq. (\ref{eq:Markov}) reduces to the 
standard on-shell Boltzmann equation 
\cite{Cassing,Blaizot:2001nr}:
\beq
\frac{\del}{\del t} N(\p,t) &=&
- \frac{\lambda^2}{2 \wm} \int \frac{d^3 k_1}{(2\pi)^3}
\frac{d^3 k_2}{(2\pi)^3} \frac{\pi \delta(\wone+\wtwo-\wmin-\wm)}
{8\wone \wtwo \wmin}
\nonumber \\
&&\times \left\{ 
\left[\bar{N}_{\k_1}\bar{N}_{\k_2}N_{\k_1+\k_2-\p} N_{\p}
- N_{\k_1}N_{\k_2}\bar{N}_{\k_1+\k_2-\p} \bar{N}_{\p}\right]
\right\}
\, .
\label{eq:onshell_Boltzmann}
\eeq
In this equation, the time integration in the original KB equation reduces to
$\delta$ function which represents the energy conservation in each
binary collisions. Therefore, the quantum effect involved in the
retardation is considered to be related to the energy broadening.
It should be noted that while only the binary collisions contribute to
the collision term in the on-shell Boltzmann equation 
because these collisions have no kinematical
threshold on the mass-shell, $0 \leftrightarrow 4$,
$1 \leftrightarrow 3$ and $2 \leftrightarrow 2$ scattering processes are
effective in the kinetic equations (\ref{eq:nonMarkov}) and (\ref{eq:Markov}).

\subsection{Single particle excitation}

The kinetic equations obtained above
are nonlinear equations for the
Wigner distribution function. As an application to practical
calculations, let us consider a single particle excitation: We add a
particle with momentum $\p$ at $t_0 = 0$ to a system which is in thermal
equilibrium initially. We would like to calculate the relaxation rate 
for such an excitation. Since the distribution functions with momenta 
$\k \neq \p$ in the collision term don't change significantly from the thermal
equilibrium value $N(\wk)$, we can take the self-energies 
(equivalently, the distribution functions with $\k \neq \p$) 
as in thermal equilibrium \cite{Blaizot:2001nr}. 
Under this situation, the kinetic
equations reduce to linear equations for 
$\delta N(\p,t) \, \left(\equiv N(\p,t) - N(\wm) \right)$ as follows:
Eq. (\ref{eq:Markov}) without the memory effect reduces to
\be
\frac{\del}{\del t} \delta N(\p,t) = -2 \gamma(\p,t) \delta
N(\p,t) \, ,
\label{eq:Markov_SPE}
\ee
and Eq. (\ref{eq:nonMarkov}) with the memory effect to
\beq
\frac{\del}{\del t} \delta N(\p,t) &=& 2 \int_0^t dz_0 \left(
\frac{\del}{\del z_0} \gamma(\p,t-z_0) \right) \delta N(\p,z_0) \nonumber \\
&=& -2 \gamma(\p,t) \delta N(\p,0) -2 \int_0^t dz_0 \gamma(\p,t-z_0)
\frac{\del}{\del z_0} \delta N(\p,z_0) \, .
\label{eq:nonMarkov_SPE}
\eeq
Here, $\gamma(\p,t)$ is a time-dependent
damping rate given by
\beq
&& \gamma(\p,t) = \frac{\lambda^2}{12 \wm} \int \frac{d^3 k_1}{(2\pi)^3}
\frac{d^3 k_2}{(2\pi)^3} \frac{1}{8\wone \wtwo \wmin}
\nonumber \\
&&\times \left\{ 
\frac{\sin(\wone+\wtwo+\wmin-\wm)t}{\wone+\wtwo+\wmin-\wm}
\left[\bar{N}(\wone) \bar{N}(\wtwo) \bar{N}(\wmin)
- N(\wone) N(\wtwo) N(\wmin) \right] \right.
\nonumber \\
&&\hspace{0.3cm}+ 
\frac{\sin(\wone+\wtwo+\wmin+\wm)t}{\wone+\wtwo+\wmin+\wm}
\left[ N(\wone) N(\wtwo) N(\wmin)
- \bar{N}(\wone) \bar{N}(\wtwo) \bar{N}(\wmin) \right]
\nonumber \\
&&\hspace{0.3cm}+ 3 
\frac{\sin(\wone+\wtwo-\wmin-\wm)t}{\wone+\wtwo-\wmin-\wm}
\left[\bar{N}(\wone) \bar{N}(\wtwo) N(\wmin)
- N(\wone)N(\wtwo)\bar{N}(\wmin) \right]
\nonumber \\
&&\hspace{0.3cm}+ 3 
\frac{\sin(\wone-\wtwo-\wmin-\wm)t}{\wone-\wtwo-\wmin-\wm}
\left. \left[\bar{N}(\wone)N(\wtwo)N(\wmin)
- N(\wone)\bar{N}(\wtwo)\bar{N}(\wmin)\right]
\right\} \, ,
\label{eq:TDDR}
\eeq
where distribution functions are in thermal equilibrium, i.e. 
$N(\wk) = 1/(\mbox{exp}(\wk/T) - 1)$ with temperature $T$, and 
$\wk$ is independent of time. 
We have used the property $\gamma(\p,t=0) = 0 $ 
from the first line to the second in (\ref{eq:nonMarkov_SPE}).
It it important that the right-hand side of
(\ref{eq:nonMarkov_SPE}) at some time $t$ can be computed by integrating
over the known functions for the time range $z_0 < t$ with given
initial value of $\delta N(\p,0)$ thanks to $\gamma(\p,0) = 0 $.
It is noted that this
time-dependent damping rate is well defined even in a theory with
long-range interactions like gauge theories in which self-energies in
the on-shell limit are ill-defined
\cite{Blaizot:2001nr}.

In order to highlight the non-Markovian nature,
it is instructive to consider the infinite-time limit $t \to \infty$
in which the time-dependent damping rate
reduces to the on-shell damping rate 
$\gamma_{\infty}(\p)$ given by
\beq
\gamma_{\infty}(\p)  =
\frac{\lambda^2}{4 \wm} &\int& \frac{d^3 k_1}{(2\pi)^3}
\frac{d^3 k_2}{(2\pi)^3} 
\frac{\pi \delta(\wone+\wtwo-\wmin-\wm)}{8\wone \wtwo \wmin}
\nonumber \\
&\times& 
\left[\bar{N}(\wone) \bar{N}(\wtwo) N(\wmin)
- N(\wone)N(\wtwo)\bar{N}(\wmin) \right]
\, .
\label{eq:onshell_dampingrate}
\eeq
This quantity has been obtained analytically for $\p = 0$
\cite{Parwani:1991gq}
and numerically for $\p \neq 0$ 
\cite{Jeon:if,Wang:1995qf}.
In this limit, (\ref{eq:Markov_SPE}) and (\ref{eq:nonMarkov_SPE})
coincide with the on-shell Boltzmann equation for a single particle
excitation,
\be
\frac{\del}{\del t} \delta N(\p,t) = -2 
\gamma_{\infty}(\p) \delta
N(\p,t) \, ,
\label{eq:onshell_SPE}
\ee
whose solution represents an exponential damping of the excitation
\be 
\delta N(\p,t) = \delta N(\p,0) \, \mbox{exp}\left(- 
\tau_{\mathrm rel}^{-1}(\p) \cdot t \right)
\label{eq:exp_damping}
\, ,
\ee
with the relaxation time 
$\tau_{\mathrm rel}(\p) \equiv 1/2\gamma_{\infty}(\p)$. 
The non-Markovian nature, or the memory effect, disappears 
in this limit. We, therefore, realize that it is related to the
short-time phenomena.

\section{Numerical results}

In this section, we numerically calculate the thermal mass and the
time-dependent damping rate for several values of the coupling constant.
Then, we solve the kinetic equations with and without the memory effect.
As expected, the relaxation time coincides that of the on-shell
Boltzmann equation in the weak coupling regime due to a separation of scales.
On the other hand, the results clearly demonstrate 
that the memory effect can influence
relaxation of the single particle excitation in the moderately large and
the strong coupling regimes.

\subsection{Renormalization of gap equation}

Since Green's functions in the self-energies are
dressed ones as mentioned in Sec. II. A and the quasiparticle mass
squared $M^2$ depends on the Green's functions, we have a  
a following gap equation for the thermal self-energies, 
\beq
M^2 &\equiv& m^2 + \Sigma^{\delta}(x_0,x_0)
\nonumber \\
&=& m^2 + \frac{\lambda}{2} \int \frac{d^3 k}{(2\pi)^3}
\frac{2 N(\wk) +1}{2 \wk} \, ,
\label{eq:unren_gap}
\eeq
where the thermal mass $M$ is independent of time, and 
$\wk = \sqrt{\k^2 + M^2}$. 
\footnote{
For the thermal (static) self-energy, $G^<$ takes a same form in the KB
and GKB ansatz, and consequently the corresponding gap 
equations are the same as (\ref{eq:unren_gap}).  
         }
After eliminating the ultraviolet divergences in the momentum
integral in the modified MS scheme, one obtains the
renormalized gap equation 
which contains no UV divergences
\cite{Blaizot:2000fc,Drummond:1997cw,vanHees},
\be
M^2 = m_r^2 + \frac{\lambda_r M^2}{32 \pi^2}\left( \ln
\frac{M^2}{\bar{\mu}^2} -1 \right) + \frac{\lambda_r}{2}\int 
\frac{d^3 k}{(2\pi)^3} \frac{N(\wk)}{\wk} \, .
\label{eq:ren_gap}
\ee
Here, $\bar{\mu}$ is an arbitrary subtraction scale, 
$m_r$ is the renormalized mass, and $\lambda_r$ denotes the
renormalized coupling constant which satisfies
\be
\frac{d \lambda_r}{d \ln \bar{\mu}} = \frac{\lambda_r^2}{16 \pi^2} \, .
\ee
This relation assures the solution $M$ of (\ref{eq:ren_gap}) does not
depend on $\bar{\mu}$. Below, we denote the renormalized mass and
coupling as $m$ and $\lambda$ for simplicity. By setting $\bar{\mu}$ 
and $m$, $M$ is obtained self-consistently for given $\lambda$ from
Eq. (\ref{eq:ren_gap}).

\begin{figure}[htbp]
    \centerline{
      \epsfxsize=0.49\textwidth
      \epsfbox{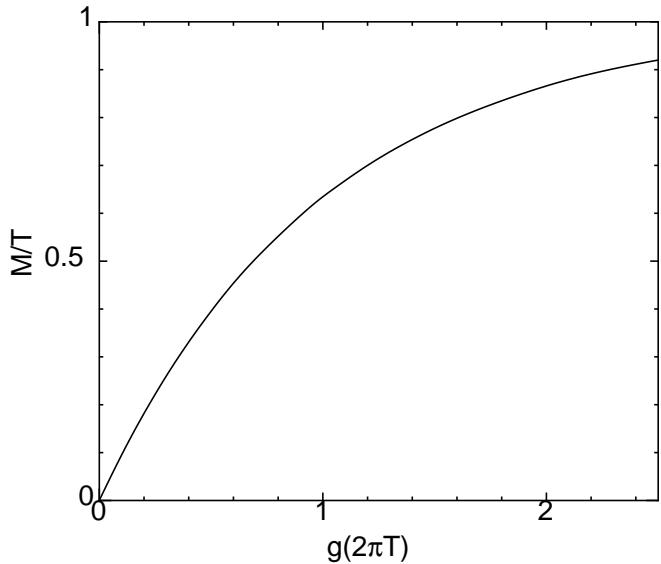}
    }
  \caption{Thermal mass $M$ over $T$ is plotted as a function of
      the coupling constant. Renormalization scale is set to $2 \pi T$.
          } 
  \label{fig:gap}
\end{figure}

Figure \ref{fig:gap} shows the numerical results of $M/T$ 
as a function of the coupling constant $g$ with 
$m = 0$ and $\bar{\mu} = 2 \pi T$.
(Here and below, the coupling constant is rescaled via
$\lambda = 24 g^2$.)
It is found that $M/T$ increases monotonically as $g$ is larger.
This result agrees with the result in \cite{Blaizot:2000fc}.
In the following calculations, 
we will use the solution $M$ obtained here.

\subsection{Time-dependent damping rate}

Next we examine the time-dependent damping rate given by (\ref{eq:TDDR}). 
Below, we focus on the zero mode $\p = 0$.
For $\p = 0$, (\ref{eq:TDDR}) reduces to 
\beq
\frac{\gamma (\p=0,t)}{(24 g^2)^2} &=& \frac{1}{12M}\frac{1}{64 \pi^4} 
\int_0^{\infty} dk_1 dk_2 \frac{k_1 k_2}{\wone \wtwo} \left. \frac{}{}\right\{
 \mbox{Si}\left((\wone + \wtwo + \omega_+ -M)\cdot t\right)
      -\mbox{Si}\left((\wone + \wtwo + \omega_- -M)\cdot t\right) 
\nonumber \\
&& \hspace{4.8cm} - 
 \mbox{Si}\left((\wone + \wtwo + \omega_+ +M)\cdot t\right)
      + \mbox{Si}\left((\wone + \wtwo + \omega_- +M)\cdot t\right)
\nonumber \\
&&+ 3 N(\wone) \left[ \frac{}{}
 \mbox{Si}\left((\wone + \wtwo + \omega_+ +M)\cdot t\right)
      -\mbox{Si}\left((\wone + \wtwo + \omega_- +M)\cdot t\right)
       \right. 
\nonumber \\
&& \hspace{1.7cm} - 
 \mbox{Si}\left((\wone + \wtwo + \omega_+ +M)\cdot t\right)
      + \mbox{Si}\left((\wone + \wtwo + \omega_- +M)\cdot t\right) 
\nonumber \\
&& \hspace{1.7cm} + 
 \mbox{Si}\left((-\wone + \wtwo + \omega_+ -M)\cdot t\right)
      -\mbox{Si}\left((-\wone + \wtwo + \omega_- -M)\cdot t\right)
\nonumber \\
&& \hspace{1.7cm} 
+\mbox{Si}\left((\wone - \wtwo - \omega_+ -M)\cdot t\right)
      -\mbox{Si}\left((\wone - \wtwo - \omega_- -M)\cdot t\right)
      \left. \frac{}{}\right]
\nonumber \\
&&+ 3 N(\wone) N(\wtwo) \left[ \frac{}{}
 \mbox{Si}\left((\wone + \wtwo + \omega_+ +M)\cdot t\right)
      -\mbox{Si}\left((\wone + \wtwo + \omega_- +M)\cdot t\right)
      \right. 
\nonumber \\
&& \hspace{2.85cm} - 
 \mbox{Si}\left((\wone + \wtwo + \omega_+ +M)\cdot t\right)
      +\mbox{Si}\left((\wone + \wtwo + \omega_- +M)\cdot t\right) 
\nonumber \\
&& \hspace{2.7cm} + 2
 \mbox{Si}\left((-\wone + \wtwo + \omega_+ -M)\cdot t\right)
      -2\mbox{Si}\left((-\wone + \wtwo + \omega_- -M)\cdot t\right) 
\nonumber \\
&& \hspace{2.7cm} + 2
\mbox{Si}\left((\wone - \wtwo - \omega_+ -M)\cdot t\right)
      -2\mbox{Si}\left((\wone - \wtwo - \omega_- -M)\cdot t\right)
\nonumber \\
&& \hspace{2.85cm} 
+\mbox{Si}\left((\wone + \wtwo - \omega_+ -M)\cdot t\right)
      -\mbox{Si}\left((\wone + \wtwo - \omega_- -M)\cdot t\right) 
\nonumber \\
&& \hspace{2.85cm} +
 \mbox{Si}\left((-\wone - \wtwo + \omega_+ -M)\cdot t\right)
      -\mbox{Si}\left((-\wone - \wtwo + \omega_- -M)\cdot t\right)
\left. \left. \frac{}{}\right] \frac{}{}\right\} \, ,
\label{eq:TDDR_zeromode}
\eeq
where $k_{1,2} \equiv |\k_{1,2}|$, 
$\omega_{\pm} \equiv \sqrt{(k_1 \pm k_2)^2 + M^2}$, and Si denotes 
the sine integral defined by $\mbox{Si}(x) \equiv \int_0^x dt \sin(t)/t$. 

\begin{figure}[htbp]
    \centerline{
      \epsfxsize=0.49\textwidth
      \epsfbox{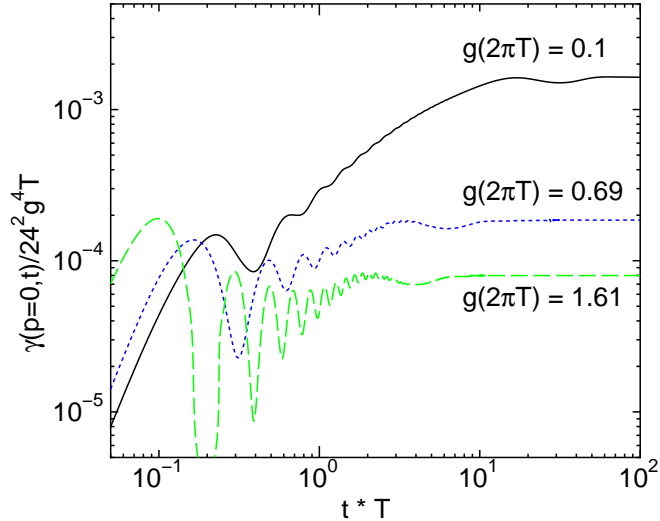}
    }
  \caption{The time dependence of $\gamma(\p=0,t)/(24g^2)^2 T$ 
           for three values of the coupling constant,
           $g(2\pi T) = 0.1\,({\mathrm solid})$, $0.69\,({\mathrm dotted})$ 
           and $1.61\,({\mathrm dashed})$.
          }
  \label{fig:TDDR}
\end{figure}

In Fig. \ref{fig:TDDR}, $\gamma(\p=0,t)/24^2g^4T$ is plotted 
as a function of $t\cdot T$ for three values of the coupling
constant $g(2\pi T) = 0.1, 0.69$ and $1.61$. As we can see,
for all values of the coupling constant,
$\gamma(\p=0,t)$ oscillates in early time and then approaches to some
constant value which corresponds to the value in the infinite-time limit, 
i.e. $\gamma_{\infty}(\p=0)$.
As the coupling constant is larger,
it is quicker for $\gamma(\p=0,t)$ to end oscillating.
\footnote{
It should be noted that if we use the bare mass $m$ 
instead of the thermal mass $M$ in the sunset self-energies, 
the time evolution of $\gamma(\p,t)/24^2g^4T$ is exactly the same
for any value of the coupling constant. 
It is important to include a mean-field effect on the propagation of
particles generated by the tadpole self-energy
as well as a scattering effect inherent in the sunset self-energies
on damping of the excitation. 
         }

\subsection{Decay of single particle excitation:
comparison between the cases with and without memory effect}

Now let us examine the time evolution of the distribution function
with/without the memory effect.
Below, we will call $\delta N(\p=0,t)$ obtained 
from (\ref{eq:Markov_SPE}) ``{\it Markovian result}'' and that from
(\ref{eq:nonMarkov_SPE}) ``{\it non-Markovian result}'', and 
make a comparison between Markovian and non-Markovian results 
for several values of the coupling constant.

\subsubsection{Weak coupling regime}

\begin{figure}[htbp]
    \centerline{
      \epsfxsize=0.49\textwidth
      \epsfbox{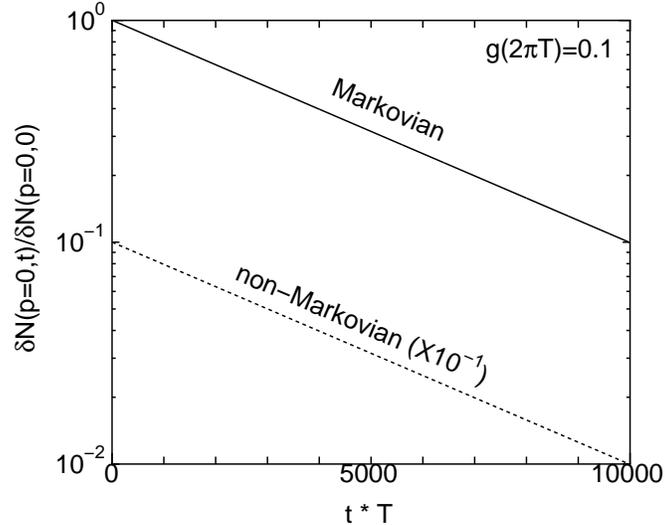}
    }
  \caption{ $\delta N(\p=0,t)/\delta N(\p=0,0)$ for Markovian and
      Non-Markovian results is plotted as a function of
      $t \cdot T$. Non-Markovian result has been divided by $10$.
          } 
  \label{fig:dn_weak}
\end{figure}

Figure \ref{fig:dn_weak} shows the time evolutions of $\delta N(\p=0,t)$ 
for the weak coupling $g(2 \pi T) = 0.1$.
As we can see, the Markovian and non-Markovian results agree completely.
This is because a separation of scales is satisfied in the weak coupling
regime: the relaxation time is much larger than the time at which the
oscillation of the time-dependent damping rate ends. 
For $g(2 \pi T) = 0.1$, the relaxation time (times $T$) is about $4300$, 
and oscillation of $\gamma(\p=0,t)$ ends at $t\cdot T \sim 20$, 
as read directly from Figs. \ref{fig:dn_weak} and \ref{fig:TDDR}. 
In this case, $\delta N(\p=0,t)$ essentially obeys the on-shell
Boltzmann equation (\ref{eq:onshell_SPE}) and
represents the exponential damping (\ref{eq:exp_damping}) in both cases. 

\subsubsection{Moderately large coupling regime}

\begin{figure}[htbp]
    \centerline{
      \epsfxsize=0.49\textwidth
      \epsfbox{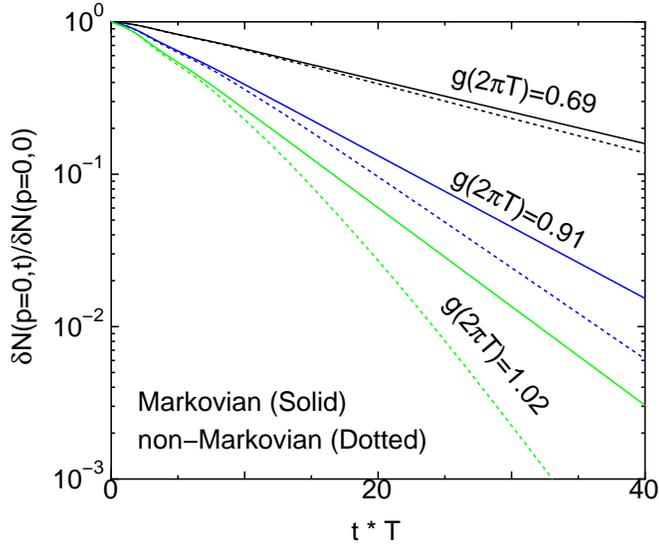}
    }
  \caption{Similar to Fig. \ref{fig:dn_weak} for the coupling constant
      $g(2 \pi T) = 0.69, 0.91$ and $1.02$ from the top to the bottom. 
      Solid and dotted lines represent Markovian and non-Markovian results. 
          } 
  \label{fig:dn_large}
\end{figure}

In Fig. \ref{fig:dn_large}, $\delta N(\p=0,t) / \delta N(\p=0,0)$
is plotted as a function of $t \cdot T$ in the moderately large
coupling regime, $g(2\pi T) = 0.69, 0.91$ and $1.02$. 
In this regime, relaxation for the non-Markovian case
is faster than that for the Markovian case, which is consistent with the
results from the KB equations in 2+1 dimensions \cite{Cassing}.

\subsubsection{Strong coupling regime:
Oscillation of the distribution function}

\begin{figure}[htbp]
    \centerline{
      \epsfxsize=0.49\textwidth
      \epsfbox{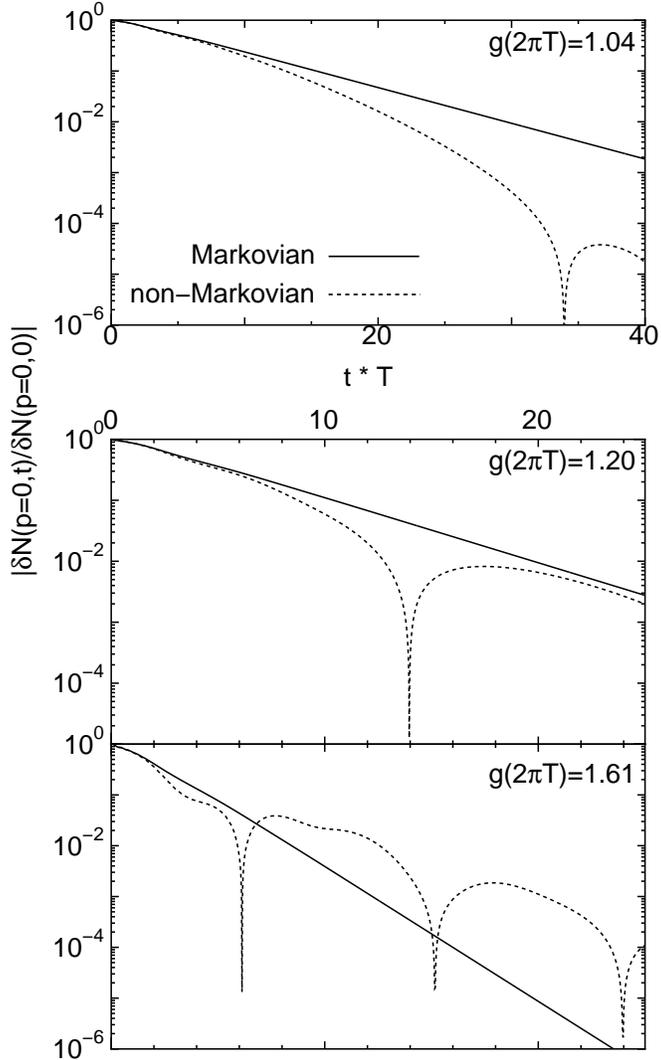}
    }
  \caption{Absolute values of $\delta N(\p=0,t)/\delta N(\p=0,0)$ 
      for Markovian (solid
      lines) and non-Markovian (dotted lines) are plotted as a function of
      $t \cdot T$ for the coupling constant $g(2\pi T) = 1.04, 1.20$ and 1.61,
      from the top to the bottom panels.
          } 
  \label{fig:dn_strong}
\end{figure}

For the strong coupling constants $g(2\pi T) = 1.04, 1.20$ and $1.61$, 
Fig. \ref{fig:dn_strong} shows absolute values of 
$\delta N(\p=0,t) / \delta N(\p=0,0)$ as a function of $t \cdot T$.
We observe that $\delta N(\p=0,t)$ for the non-Markovian case
oscillates around zero for $g \ge 1.04$.
\footnote{We have checked that $\delta N$ with the memory effect 
doesn't oscillate for $g < 1.04$ by seeing the behaviors of 
$\delta N$ and $\del_t \delta N$ in the late time. 
For $g < 1.04$, as time evolves, $\delta N /\delta N(t=0)$ starting from
unity and $- \del_t \delta N /\delta N(t=0)$ from zero become parallel
and never cross. They decrease monotonically in the late time, and
$\delta N$ cannot reach the negative value. 
         }
As we can see in the middle panel, 
relaxation for Markovian and non-Markovian cases equals at 
$g \sim 1.2$. Then, relaxation for the non-Markovian case 
becomes slower than that for the Markovian case,
as read from the bottom panel.

\subsection{Discussion}

Only in the weak coupling regime where damping of
the excitation is exponential, the relaxation time $\tau_{\mathrm{rel}}$
can be obtained by an exponential fitting
\be
\delta N(\p,t) \sim \exp\left( -\tau_{\mathrm{rel}}^{-1} 
\cdot t \right) \, .
\ee
In the moderately large and the strong coupling regimes, especially 
in the region where $\delta N$ oscillates, we can not define 
the relaxation time of the zero mode in a precise sense. 
The important point, however, is that 
an opposing influence of the memory effect
on relaxation of the excitation is observed as the 
coupling constant changes, as you can see in Figs. \ref{fig:dn_large} 
and \ref{fig:dn_strong}.

Referring to the discussion in the weak coupling regime,
it should be noted that the memory effect can
affect the relaxation time even in the region where the separation of
scales is satisfied well. For $g = 0.69$,
although the relaxation time ($\sim 23 \cdot T$) in Fig. \ref{fig:dn_large} is
larger than $\sim 3 \cdot T$ at which the oscillation of $\gamma(\p,t)$
ends in Fig. \ref{fig:TDDR}, there is the onset of the memory effect,
that is, there is a slight difference between Markovian and
non-Markovian results.
Therefore, it is expected that the separation of scales works only in the
adequately weak coupling.

In Fig. \ref{fig:dn_strong}, as the coupling constant increases,
the relaxation time for the non-Markovian results seems 
inclined to saturate.
On the other hand, the Markovian results show a monotone decreasing
of the relaxation time. 
In order to see whether the relaxation time saturates, we need to 
study the time evolution for stronger coupling $g > 1.6$, which 
is computable in the same way. However, such a strong coupling
brings about the possibility of conflicting with the quasiparticle
approximation used in our calculations, as mentioned below. 
Therefore, we don't make further consideration for 
the possibility of a saturation of the relaxation
time here, which is an important future work.

Throughout the present paper,
we have used the quasiparticle approximation in which 
an excitation has an infinite lifetime. The validity of this
approximation requires the condition that the relaxation time is larger
than $1/\wm$ ($= 1/M$ for zero mode), 
which corresponds to a long-lived excitation
\cite{Blaizot:2001nr}. 
In the range of the coupling constant in our calculations, 
this condition are satisfied. 
In order to examine the case of stronger coupling constant, it is
needed to include the dynamical spectral function with the finite
width, which is outside the present scope.

\section{Summary and outlook}

Based on the kinetic equations with/without the memory effect, 
we studied the role of that effect in decay of the single particle
excitation of $\phi^4$ theory. 
Starting from the two-particle irreducible effective action 
of the three-loop approximation and 
corresponding Kadanoff-Baym equations,
we derived the kinetic equations into which 
the memory effect was
incorporated by the GKB ansatz 
proposed by P.\ Lipavsky, V.\ Spicka, and B.\ Velicky 
\cite{Lipavsky}.
It was observed in numerical results that, while relaxation of the
excitation is unaffected by the memory effect in the weak coupling regime, 
this effect influences it
differently as the coupling constant increases: The memory effect 
leads to faster relaxation for the moderately large coupling
and to slower relaxation for $g > 1.2$.
It is expected that the effect of memory
manifests itself in totally nonequilibrium situation, that is,
in non-linear kinetic equations for the distribution functions. 
The important point is that, although we adopted a simple $\phi^4$
model, the influence of the memory effect on relaxation 
is probably universal in strongly correlated plasmas.
In particular, our results suggest a possibility of the saturation 
of the relaxation time.

Finally, some comments are in order. The spectral function used in this
paper is the quasiparticle one and cannot be applicable to the system 
where the relaxation time is shorter than $1/\wm$. In order to study the
effect of the finite width in the spectral function, it would be
necessary to use the phenomenological treatment such as an extended
quasiparticle picture \cite{Kohler2,Spicka}, 
or to solve directly KB equations for two-point Green's functions
\cite{Aarts:2001qa,Berges:2000ur,Berges:2001fi,Berges,Berges:2002wr,Cassing}. 
Two-time Green's functions in KB equations involve the information 
about the dynamical spectral function having the finite width as well as
the one-particle distribution function.
Furthermore, study of the linear sigma model or
the fermionic system with the chiral phase transition is also an
important future project in our approach, which might be related to
Heavy-ion collision experiments.

\section*{ACKNOWLEDGMENTS}

I am grateful to Y.~Nemoto for numerous valuable discussions. 
This work was supported by Special Postdoctoral Researchers Program 
of RIKEN. 


%
 
%

%
%

\end{document}